\documentclass{aa}
\usepackage{graphicx}
\begin{document}
   \title{Temperature and gravity of the pulsating
extreme helium star \object{LSS~3184} (\object{BX~Cir}) 
through its pulsation cycle
\thanks{Based on observations obtained at the
Anglo-Australian Telescope, Coonabarabran, NSW, Australia.}}

   \author{Vincent M. Woolf \thanks{Current address: Astronomy Department,
University of Washington, Box 351580, Seattle, WA 98195-1580, USA}
          \and
          C. Simon Jeffery 
          }

   \offprints{V. M. Woolf}

   \institute{Armagh Observatory, College Hill, Armagh, BT61~9DG, Northern
Ireland \\
              \email{vmw@astro.washington.edu,csj@star.arm.ac.uk}
             }

   \date{}

   \authorrunning{V.M. Woolf \& C.S. Jeffery}
   \titlerunning{Temperature and gravity of LSS~3184}     

   \abstract{

We report the analysis of optical spectra of the extreme helium star
LSS~3184 (BX~Cir) to determine its effective temperature and gravity
throughout its pulsation cycle. The spectra were also used to measure its
chemical abundances.

We report rest gravity, $\log g = 3.38 \pm 0.02$, and a chemical abundance
mixture consistent with those reported earlier in a study using an optical
spectrum with lower spectral resolution and a lower signal to noise ratio.
Our analysis decreases the upper limit for the H abundance to ${\rm H < 6.0}$
(mass fraction $< 7.1 \times 10^{-7} $).  Our gravity corresponds to stellar
mass $M = 0.47 \pm 0.03 M_\odot$.

We find that the effective $\log g$ varies through the pulsation cycle with an
amplitude of 0.28~dex.  The effective gravity is smaller than the rest
gravity except when the star is very near its minimum radius.  The change in
effective gravity is primarily caused by acceleration of the stellar
surface.

Based on the optical spectra, we find the temperature varies with an amplitude
of 3450~K.
We find a time averaged mean temperature, $23390 \pm 90$~K, consistent with
that found in the earlier optical spectrum study.  The mean temperature is
1750~K hotter than that found using combined ultraviolet spectra and V and R
photometry and the variation amplitude is larger.  This discrepancy
is similar to that found for the extreme helium star \object{V652~Her}.

      \keywords{stars: chemically peculiar -- stars: oscillations --
stars: variables -- stars: individual: LSS~3184 -- stars: atmospheres
               }  
   }

   \maketitle
%

\section{Introduction}

Extreme Helium stars (EHes) are a class of low-gravity, hot, evolved stars
with very large helium abundances ($\ga 99$ per cent) and weak or non-existent
hydrogen lines.  They appear to be rapidly evolving to become white dwarfs.
The short time the stars spend as EHes explains why few EHes are known, despite
their brightness ($L \ga 900 L_\odot$).

The evolutionary history of EHes is still uncertain. The more popular
proposals have been that they are formed by the merger of two white
dwarfs (Webbink \cite{w84}, Iben \& Tutukov \cite{it84}) or that they are the
result of a `late thermal pulse' when the helium shell near the surface of a
white dwarf ignites, causing the star to expand (Iben et al. \cite{i83}).
Saio \& Jeffery (\cite{sj02}) have recently shown that the white dwarf
merger model is the most likely explanation.  Determining
the history of EHes will provide clues to the evolutionary history of possibly
related stars (RCrB stars, He-rich subdwarfs, carbon stars, etc.).  As the
chemical abundance mixture in EHes appears to be the end result of a
combination of CNO and triple-$\alpha$ processing,
their study may provide information about physical processes occurring in a
large fraction of stars.

Some EHes pulsate. The pulsations are driven through the $\kappa$ mechanism,
with iron group (Z-bump) elements providing the needed opacity
(Saio \cite{s95}).
Pulsation provides additional methods to study the physical properties of stars.
Radial velocities, ultraviolet spectra, and optical photometry
have been used to determine the radius, gravity, temperature, mass, and
absolute magnitude of two pulsating EHes: LSS~3184 and
V652~Her (Lynas-Gray et~al. \cite{lg84}, Kilkenny et al. \cite{k99}, Woolf
\& Jeffery \cite{wj00}, Jeffery et~al. \cite{jwp01}).
Mass and chemical composition
are the parameters which best constrain evolutionary models for EHes.
In previous studies of LSS~3184 the relatively
large uncertainty for the gravity produced most of the uncertainty for the
mass determination.

In a recent study of pulsation models for EHes, Monta\~n\'ez Rodr\'{\i}guez \&
Jeffery (\cite{mj02}) showed that the pulsation
period and radial velocity curve of LSS~3184 require a mass between 0.38 and
0.5 $M_\odot$ if the temperature is assumed to be between 22\,400 and 24\,000~K.
However, the models with those parameters also require a luminosity smaller than
half of that previously measured.

In this paper we report the determination of chemical abundances and time
resolved temperature and effective gravity of LSS~3184 using
spectral analysis of the optical spectra used to find radial velocities in
Woolf \& Jeffery (\cite{wj00}).


\section{Observations and data reduction}
The optical spectra used in this study are those used to determine radial
velocities through the pulsation cycle of LSS~3184 as
described in Woolf \& Jeffery (\cite{wj00}). The spectra were obtained at the
Anglo-Australian Telescope on the nights of 1996 May 18 and 19
using the University College London Echelle Spectrograph.  Standard
{\sc iraf} packages were used for bias and flat field correction, reducing
echelle orders to one dimensional spectra, and applying the wavelength scale
using thorium-argon arc spectra.  The spectral resolution is
$\lambda / \Delta \lambda \approx 48\,000$.

We merged the echelle orders to produce a single spectrum for each observation
covering the entire observed wavelength range.  This is a fairly complicated
procedure because the fluxes rarely match and frequently have opposite slopes
in the sections of the echelle orders which overlap in wavelength. This is
caused mostly by incomplete correction for the grating blaze function.

Combining the echelle orders without properly correcting this produces
discontinuities at the wavelengths at which echelle orders begin and end.
Normalizing the echelle orders by dividing by estimates of the continuum
levels before merging reduces the size of the discontinuities but
does not eliminate them.  Furthermore, because there are uncertainties in
continuum fitting, using this method makes it difficult to
determine if a difference between two spectra is caused by stellar
variability or by an inconsistency in continuum fitting. This problem is worse
when a strong absorption feature appears near or at the end of an echelle order.

To overcome this problem, we coadded the spectra to produce an average spectrum
with a high signal to noise ratio.  Continuum shapes were estimated for each
echelle order where it was reasonable to do so, i.e. orders without
strong absorption features.  Second degree polynomials, $y = a+bx+cx^2$,  
were sufficient to reproduce the curvatures present.  Fifth degree polynomials
were then fitted to the coefficient ($a, b, c$) vs echelle order curves and
used to calculate ``smoothed'' continuum fit coefficients for all orders,
including those with strong absorption features.

The smoothed continuum fits calculated with the average
spectrum were then used to
normalize individual spectra.  The {\sc iraf} procedure {\sc scombine} was
used to combine the echelle orders into a single normalized spectrum without
the problem of discontinuities previously present.  

This problem has been treated by others using somewhat different methods
(e.g. Barker \cite{b84}, Erspamer \& North \cite{en02}).  

\section{Analysis}
We determined that the individual spectra were too noisy to be used to find
the temperature, gravity, and chemical composition of LSS~3184
reliably.  We combined the spectra into twelve phase bins to provide spectra
with high enough signal to noise ratios (STN) for our analysis
$(30 < STN < 50)$ while also providing
small enough time increments to allow us to measure changes to the atmospheric
conditions through the pulsation cycle. The bins were selected so that the
second one is centered at minimum
radius (phase = 0.1).  Bins are one twelfth of a cycle long ($P = 0.1065784$~d)
(Kilkenny et al. \cite{k99}) and contain between 10 and 15 individual spectra.

Determining microturbulent velocity was an iterative process.  C and O
abundances were found using absorption line equivalent widths and
estimates of $\log g$ and $T_{\rm eff}$ in the program {\sc spectrum}
(Jeffery et~al. \cite{jwp01}).  Microturbulence was varied until
there was no slope in the C and O abundance vs equivalent width plots. We
calculated $\log g$ and $T_{\rm eff}$ assuming this microturbulence (Sect.
3.1). We then tested that the previously determined
microturbulence produced C and O abundance vs equivalent width
plots with no slope with the newly determined $\log g$ and $T_{\rm eff}$.

We found $v \sin i$ by fitting line widths in synthetic spectra to those in
the optical spectra.  Other line-broadening effects which do not affect
line saturation such as macroturbulence and pulsation velocity change during
the exposure may be included in the number we found. Based on the accelerations
measured in Woolf \& Jeffery (\cite{wj00}), the surface velocity change during
a 5 minute exposure is about ${\rm 1\, km s^{-1}}$ for most of the pulsation
cycle and about ${\rm 12\, km s^{-1}}$ near minimum radius.  Including this
velocity drift in our analysis had no substantial appreciable effect on the
temperatures or gravities we calculated.

\subsection{Temperature and gravity}
We determined the temperature and effective gravity of LSS~3184 for each phase
bin using the {\sc fortran90} program {\sc sfit} (Jeffery et~al.
\cite{jwp01}).  In {\sc sfit} interpolated synthetic spectra are fitted to
observed spectra to find the best $T_{\rm eff}$ and $\log g$ values.
The input spectra were calculated for a fixed chemical composition
and microturbulent velocity $\xi = 6.9 \,{\rm km\,s^{-1}}$.

\begin{figure}
\resizebox{\hsize}{!}{\rotatebox{270}{\includegraphics{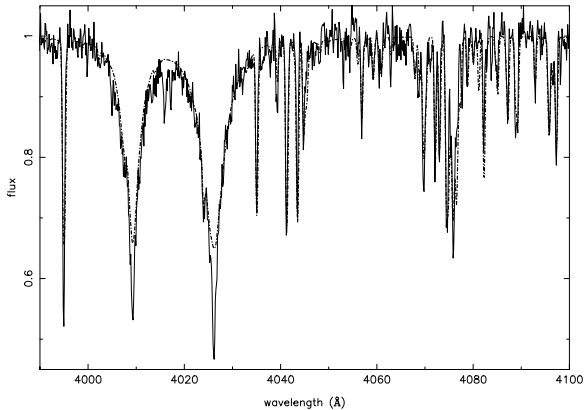}}}
\caption{Observed (solid line) and synthetic (dotted line) spectra of He
lines at 4009 and 4026~\AA\ and some metal lines. The observed spectrum is
for phase 0.6 (bin 8).  The synthetic spectrum was calculated for
$T = 21750$, $\log g = 3.08$.}
\label{fig1}
\end{figure}

\begin{figure}
\resizebox{\hsize}{!}{\rotatebox{270}{\includegraphics{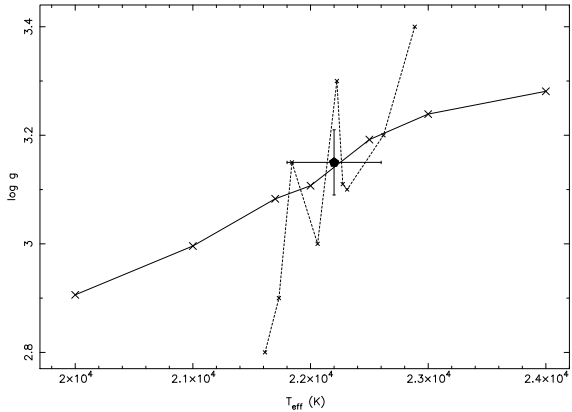}}}
\caption{$T_{\rm eff}$ found using fixed $\log g$ (dotted line) and $\log g$
found using fixed $T_{\rm eff}$ (solid line) for bin 10
(phase: $0.775<\phi<0.80833$). The large data point shows the adopted
$T_{\rm eff}$ and $\log g$ values and uncertainties for the bin
values.}
\label{fig2}
\end{figure}

Finding the best $T_{\rm eff}$ and $\log g$ values for a phase bin was a
two step process. In both steps we used $v\sin i = 15.8\,{\rm km\,s^{-1}}$.
In the first step we let $T_{\rm eff}$ vary while holding $\log g$ fixed and 
used the wavelength regions without strong
helium lines. We found the $T_{\rm eff}$ values which produced
the best fits for various input $\log g$ values.  In the second step we set 
{\sc sfit} to let $\log g$ vary while holding $T_{\rm eff}$ fixed and to use
the wavelength regions with strong helium lines. We then found the $\log g$
values which produced the best fits for various input $T_{\rm eff}$ values.
Examples of spectral fits for He lines and metal lines are shown in
Fig. \ref{fig1}.  (We note that modeling the cores of strong
He lines in hot He stars is a continuing problem.)
We estimated the temperature and gravity for each phase bin by plotting the 
$T_{\rm eff}$ vs $\log g$ data for the two steps and finding the
intersection of the two sets (Fig.~\ref{fig2}).  Uncertainty in the spectral
data and the analysis caused intersections to occur over a
range of $T_{\rm eff}$ and $\log g$.  We can therefore only determine that the
best estimate lies within the intersection range.  We believe that a
large part of this uncertainty occurs because
the spectral fitting routine adjustments to
the continuum level of the observed spectrum can improperly cancel the effects
of $\log g$ differences.

\begin{figure}
\resizebox{\hsize}{!}{\includegraphics{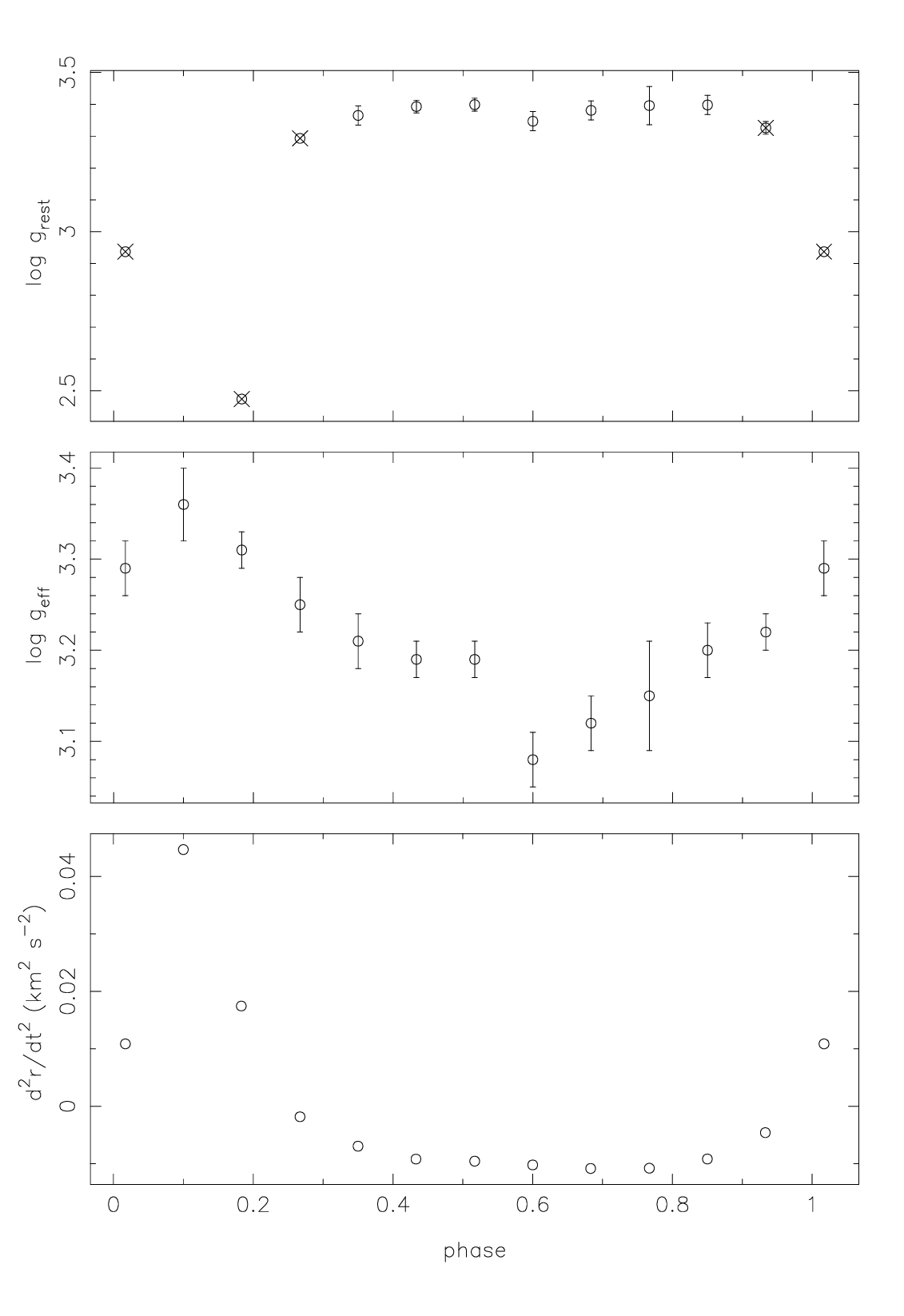}}
\caption{Linear surface acceleration (bottom panel), effective gravity (middle
panel), and gravity (top panel) of LSS~3184 through its pulsation cycle. The
$\log g_{\rm rest}$ panel does not have a point at phase $= 0$ because radial acceleration
there is larger than the measured effective gravity, giving a negative
gravity.  }
\label{fig3} 
\end{figure}

The $\log g$ values we measured are {\it effective} $\log g$ values: 
$\log g_{\rm eff} = \log (g_{\rm rest} + a)$ where $\log g_{\rm rest}$ is the
gravitational acceleration, $GM r^{-2}$, which would be measured for the
star at a current radius if it were not pulsating,
and $a$ is the ${\rm d}^2r / {\rm d}t^2$
acceleration of the surface of the pulsating star.
(To avoid confusion, hereafter we will refer to $\log g_{\rm rest}$
as ``gravity'' and ${\rm d}^2r / {\rm d}t^2$ as ``acceleration''.)
The gravity for each phase bin can be determined by subtracting the
acceleration from the measured $\log g_{\rm eff}$.  Fig.~\ref{fig3}
displays $\log g_{\rm eff}$, ${\rm d}^2r / {\rm d}t^2$, and $\log g_{\rm rest}$
through the pulsation cycle. The phase bin at
maximum acceleration (minimum radius) does not have a data point
in the upper panel because the radial acceleration
is larger than the effective gravity measured, giving a negative gravity.
The points in the upper panel with X's were not used 
to calculate the average $\log g_{\rm rest}$: they appear to be affected by the
atmosphere being out of equilibrium as a result of the strong compression
caused by radial acceleration near minimum radius.
The error bars indicate formal measurement errors. They do not include 
systematic errors caused by uncertainties in the assumptions used or 
incomplete physics in our analysis.
As will be discussed in the results section,
there is obviously some important physics omitted in our analysis of the star
near minimum radius.

\subsection{Chemical composition}
The chemical composition of LSS~3184 should not change with pulsation phase.
We calculated elemental abundances using the
combined spectra from the four phase bins closest to maximum radius
(minimum temperature) when the atmosphere is changing more slowly and where
assuming local thermodynamic equilibrium conditions is more reasonable.
The spectral wavelengths were corrected to zero velocity prior to co-adding. 
This provided a representative spectrum with a high signal to noise
ratio ($\rm STN \sim 100$) for abundance analysis.

A model atmosphere was calculated for the mean temperature and gravity found
previously for the four bins.
We calculated atomic abundances using this model atmosphere
and the {\sc fortran90} program {\sc sfit\_synth} (Jeffery et al.
\cite{jwp01}) to fit synthetic spectra to the combined observed spectrum.
Fe was an exception.  With only three Fe~III lines (4139.4, 4164.8, and
4419.6~\AA ) unblended and strong enough for analysis we used the program
{\sc spectrum} to find the Fe abundance based on their equivalent widths.
We did not find any Mg lines of adequate quality to determine the Mg abundance.

\section{Results: physical properties of LSS~3184}
The effective temperature and effective gravity for each phase bin are listed
in Table~1. The uncertainties quoted for our data
are formal errors based on how
precisely the models predict temperature and gravity.  They do not include
uncertainties in the physics that went into the models.  Near minimum
radius where non-equilibrium effects occur the model uncertainties are large.

The variation of effective gravity with phase is displayed in
Fig.~\ref{fig3}.  The difference in radius at maximum and minimum provides a
gravity change of only 0.04~dex in $\log g$, so the effective gravity variation
is primarily caused by acceleration of the star's surface due to pulsation. 
When we use spectroscopy to measure gravity, what we are actually measuring is
the pressure in the atmosphere as manifested in the wings of strong lines
(He lines in EHes).  Acceleration of the atmosphere produces
corresponding pressure changes which are small when the acceleration is
small, so that the assumption of hydrostatic equilibrium in the model
atmosphere is reasonable. But at minimum radius and maximum acceleration
the measured gravity does not increase as much as expected and we calculate
a negative gravity. 
This is probably because one of the assumptions used in determining the gravity,
local equilibrium, is clearly not valid when the atmosphere is undergoing a
strong, fast acceleration and compression.  
We believe that the gravity values found at minimum radius and
nearby phase bins (missing or marked with X's in the top panel of
Fig.~\ref{fig3}) are inaccurate because of these effects.

\begin{figure}
\resizebox{\hsize}{!}{\rotatebox{270}{\includegraphics{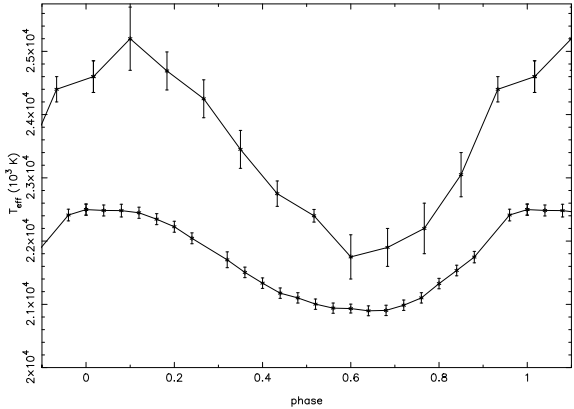}}}
\caption{$T_{\rm eff}$ vs phase for LSS~3184. The upper curve displays the
results from this paper.  The lower curve displays results of UV and visible
flux fitting in Woolf \& Jeffery (\cite{wj00}).}
\label{fig4}
\end{figure}

The temperature variation is compared to the variation found by fitting UV
and visible flux levels (Woolf \& Jeffery \cite{wj00}) in Fig.~\ref{fig4}.
The temperatures found using spectral analysis of visual spectra are hotter
by more than 1600~K
than those found using visual and UV fluxes for both LSS~3184 and V652~Her.
The temperature variation amplitude based on visible spectra is also larger.
This problem is discussed for V652~Her in Jeffery et~al. (\cite{jwp01}).

The temperature curve found by Kilkenny et al. (\cite{k99}) using optical
photometry is very similar to the one we find using optical spectral analysis.
Monta\~n\'ez Rodr\'{\i}guez \& Jeffery (\cite{mj02}) found that models which
fitted the velocity and flux curves of LSS~3184 well required a mass between
0.38 and $0.50 M_\odot$ and a temperature between 22\,300 and 23\,900~K.
Because analysis of optical spectra, optical photometry, and theoretical
models each predict the hotter temperature, it appears that the lower
temperature estimate found using UV and visual fluxes is probably in error.
That a nearly identical discrepancy was found for V652~Her means that this is
probably a systematic problem. We suspect that incomplete line opacity in our
models is the main cause.

\begin{table}
\caption{Temperature and gravity through the pulsation cycle of LSS~3184. Phases
are specified at center of phase bins.}
\begin{tabular}{lllll}
\hline
bin & phase & $T_{\rm eff}$ (K) & $\log g_{\rm eff}$ & $\log g_{\rm rest}$ \\
\hline
1 & 0.0167 & $ 24600 \pm 250 $&$ 3.29 \pm 0.03 $& 2.94 \\
2  & 0.1000 & $ 25200 \pm 500 $&$ 3.36 \pm 0.04 $& \\
3  & 0.1833 & $ 24690 \pm 300 $&$ 3.31 \pm 0.02 $& 2.47\\
4  & 0.2667 & $ 24250 \pm 300 $&$ 3.25 \pm 0.03 $& 3.29 \\
5  & 0.2500 & $ 23450 \pm 300 $&$ 3.21 \pm 0.03 $& 3.36 \\
6  & 0.4333 & $ 22750 \pm 200 $&$ 3.19 \pm 0.02 $& 3.39 \\
7  & 0.5167 & $ 22400 \pm 100 $&$ 3.19 \pm 0.02 $& 3.40 \\
8  & 0.6000 & $ 21750 \pm 350 $&$ 3.08 \pm 0.03 $& 3.35 \\
9  & 0.6833 & $ 21900 \pm 300 $&$ 3.12 \pm 0.03 $& 3.38 \\
10 & 0.7667 & $ 22200 \pm 400 $&$ 3.15 \pm 0.06 $& 3.40 \\
11 & 0.8500 & $ 23050 \pm 350 $&$ 3.20 \pm 0.03 $& 3.40 \\
12 & 0.9333 & $ 24400 \pm 200 $&$ 3.22 \pm 0.02 $& 3.33 \\
\hline
\end{tabular}
\end{table}

The physical parameters and chemical abundances derived for LSS~3184 are
presented in Table~2 and are compared with those found in previous
studies and with those of the \object{Sun}, ``typical'' hot EHes, and V652~Her.
V652~Her is a hot EHe with pulsation period, temperature, and radius nearly
identical to those of LSS~3184.
The errors quoted {\rm{\rm  for our data} are formal measurement errors and do
not account for systematic errors due to uncertainties in our input assumptions
or incomplete physics in our calculations.  The parameters we report are
consistent with those found previously using optical spectra with higher noise,
lower spectral resolution, and
no time resolution.  We found a larger gravity, though it is within the
uncertainty quoted for the previous determination.  The new gravity combined
with the stellar radius found in Woolf \& Jeffery (\cite{wj00}) produces a
mass estimate 12 per cent larger than that found using the previously
determined $\log g$.

\begin{table*}
  \begin{minipage}{115mm}
\caption{Physical parameters of LSS~3184 compared with previous determinations
and with V652~Her, Sun and average hot extreme helium
stars. Abundances use normalization: $\log \sum_i \mu_i n_i = 12.15$. 
Uncertainties listed are formal errors and do not take into account
uncertainties of input parameters or model assumptions.}
 \begin{tabular}{lllllll}
\hline
  & LSS 3184 & LSS 3184 (WJ) & LSS 3184 (KD) & V652 Her & hot EHe & Sun \cr \hline
Period (days) & & & 0.1065784 & 0.10799319 \\
$\langle T_{\rm eff} \rangle$ (K)& $23390 \pm 90$ & 21640 & $23300 \pm 700$ &
22930\footnote[3]{or 20950~K using UV and visible flux }&
$>13000$  \\
$\langle \log g \rangle$ & $ 3.38 \pm 0.02 $ & & $3.35 \pm 0.10$ & 3.7 \\
$\langle R \rangle (R_\odot)$ & & $2.31 \pm 0.10$ & $1.35 \pm 0.15$ & $2.31 \pm 0.02$  \\
mass ($M_\odot$) & $0.47 \pm 0.03$ & $0.42 \pm 0.12$ & 0.12 to 0.42 & $0.59 \pm 0.19$  \\
$\langle L \rangle (L_\odot)$&$ 1432 \pm 88$ & $1400 \pm 300$ & $485 \pm 135$ &$919 \pm 14$ \\

H  & $<6.0$ && $<7.72$ & 9.5 & 8.0& 12.00 \\
He & 11.54 && 11.54 & 11.54 & 11.54 & 10.99  \cr
C  & $ 8.89 \pm 0.16$& & 9.02 & 7.03 &9.3& 8.58 \cr
N  & $ 8.43 \pm 0.13$&& 8.26 & 8.9  &8.3& 8.05 \cr
O  & $ 7.87 \pm 0.10$&& 8.05 & 7.9 &8.6& 8.93 \\
Mg &                 && 7.17 & 8.1 & 7.6 & 7.58 \\
Al & $ 5.91 \pm 0.07$&& 6.04 & 6.7 & 6.1 & 6.47 \\
Si & $ 6.59 \pm 0.11$&& 6.91 & 7.7 & 7.4 & 7.55 \\
P  & $ 4.86 \pm 0.08$&& 4.96 & 5.8 & 5.7 & 5.45 \\
S  & $ 6.59 \pm 0.11$&& 6.67 & 7.4 & 7.1 & 7.33 \\
Fe & $ 6.72 \pm 0.13$&& 6.52 & 7.4 & 5.8 & 7.50 \\

\hline
\end{tabular}

Table references:

WJ LSS~3184 data: Woolf \& Jeffery \cite{wj00} (UV and visible flux)

KD LSS~3184 data: Kilkenny et al. \cite{k99}, Drilling et~al. \cite{d98} 

V652~Her: Lynas-Gray et~al. \cite{lg84}, Jeffery et~al. \cite{j86}, Jeffery
\cite{j96}, Jeffery et~al. \cite{jwp01}

Sun: Grevesse N., Noels A., \& Sauval A.J. \cite{g96}

average hot EHe:  Pandey et al. \cite{p01}

\end{minipage}
\end{table*}

The elemental abundances found for LSS~3184, V652~Her, and typical hot EHes
are shown in Fig.~\ref{fig5}. The Mg abundance for LSS~3184 is taken
from Drilling et~al. (\cite{d98}).
\begin{figure}
\resizebox{\hsize}{!}{\rotatebox{270}{\includegraphics{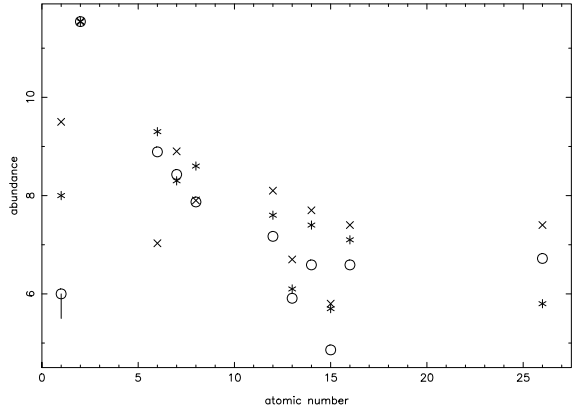}}} 
\caption{Chemical abundances of LSS~3184 (circles), V652~Her (crosses),
and typical hot EHes (asterisks) vs atomic number. The H abundance for
 LSS~3184 is an upper limit.}
\label{fig5}
\end{figure}
In the figure we can see that LSS~3184 has a much lower H abundance and
V652~Her has a much lower C abundance than the others. The H abundance we
report for LSS~3184 is an upper limit: we could find no evidence for any H
absorption lines in our spectra. As an example, the observed spectrum is
compared to synthetic spectra with different H abundances in the spectral
region including the Balmer H$\beta$ line in Fig.~\ref{fig6}. The weak
feature at 4860.97~\AA\ is an O~II line and is fitted well with no H
contribution.  The H line strengths become negligible for ${\rm H < 6.0}$
(mass fraction less than one part in $1.4 \times 10^6$).
\begin{figure}
\resizebox{\hsize}{!}{\rotatebox{270}{\includegraphics{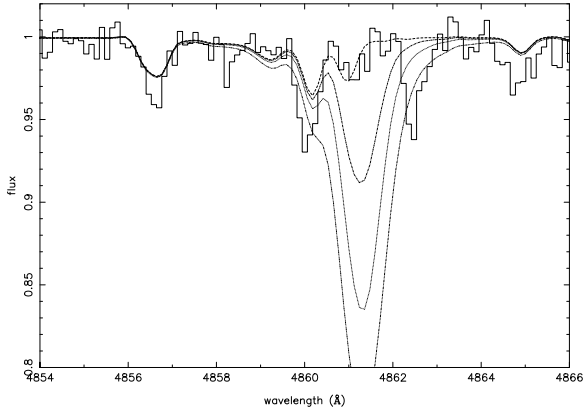}}}
\caption{ H-$\beta$ region of observed LSS~3184 spectrum and synthetic spectra
with H = 4.0, 6.0, 8.0, 8.5, and 9.0.  H = 4.0 and 6.0 are identical.  }
\label{fig6}
\end{figure} 

It is also evident from Fig.~\ref{fig5} that LSS~3184 has a
lower metallicity than V652~Her and typical hot EHes. It has a lower abundance
of all metals measured except that V652~Her has a smaller C abundance, probably
as the result of CNO processing its C to N, and the typical EHe Fe abundance is
smaller. We believe the metallicity differences indicate differences in
metallicity of the progenitor stars.

Saio \& Jeffery (\cite{sj00}, \cite{sj02})
showed that the white dwarf (WD) merger model
can explain the chemical abundances observed in EHes.  V652~Her shows evidence
of the CNO process converting essentially all carbon and oxygen into nitrogen.
LSS~3184 and typical hot EHes show evidence of CNO processing with
high N abundances, but also show evidence of the
triple-$\alpha$ process adding new C. Saio
\& Jeffery say this this could be because V652~Her is the result of a
He-He WD merger while LSS~3184 and typical hot EHes are the result of
He-CO WD mergers. A He-He WD merger for V652~Her would  also help explain
other differences (e.g. the luminosity) between it and other EHes,
including LSS~3184.

Although the gravity we find for for LSS~3184 is larger than that previously
determined, the mass to which it corresponds, $M = 0.47 \pm 0.03 M_\odot$, is
still smaller than that permitted by WD merger models for EHe production. 
It is unclear whether improved merger models or gravities found with
an improved model atmosphere
program will remove this discrepancy or if this will rule out the WD
merger model.

The hydrodynamic models of LSS~3184 which best match
its pulsation period and radial velocity curve for temperatures between
22\,400 and 24\,000 require a mass between 0.38 and $0.5 M_\odot$
(Monta\~n\'ez Rodr\'{\i}guez \& Jeffery \cite{mj02}).  The mass we find here
using optical spectral analysis ($0.47 \pm 0.03 M_\odot$) and the mass we found
by fitting synthetic spectral fluxes to UV spectra and optical photometry
($0.42 \pm 0.12 M_\odot$) Woolf \& Jeffery (\cite{wj00}) are comfortably
within this range. 

However, the Monta\~n\'ez Rodr\'{\i}guez \& Jeffery models
also predict a luminosity $2.74 < \log L/L_\odot < 2.84$
smaller than half of the luminosity found using the temperature and
radius from Woolf \& Jeffery (\cite{wj00}), $\log L/L_\odot = 3.15 \pm 0.09$ .
We do not derive a new luminosity in this paper.  Kilkenny et al.  (\cite{k99})
found $2.54 < \log L/L_\odot < 2.79$ using optical photometry.  Most of the
difference between the luminosities are caused by the differences in 
stellar radius used in the calculations: $1.519 < R/R_\odot < 1.708$ from the
pulsation models, $R = 1.35 \pm 0.15 R_\odot$ from optical photometry, and
$R = 2.35 \pm 0.10 R_\odot$ from UV plus optical flux fitting.

We suspect that
the match between the model luminosity and the optical photometry luminosity
and the mismatch with the luminosity based on UV plus optical flux fitting
is another indication that there is a problem with the part of our
code that deals with UV flux.


\section{Conclusions}
We have used time resolved spectra with high signal to noise ratios and high
spectral resolution to determine the temperature, gravity, and chemical
composition of LSS~3184.  Our results are consistent with those of previous
analyses based on optical spectra and photometry. We
have reduced the uncertainties on some of these quantities.

We have reduced the upper limit on the hydrogen abundance by a factor of 52
to $\rm H < 6.0$.  Our gravity and mass estimates are 12 per cent larger than 
previous estimates.

The time averaged mean temperature estimate we obtain with our spectral
analysis is 1750~K
hotter than that obtained by fitting UV and visible flux levels.  A similar
temperature discrepancy was found for V652~Her.  The problem is probably
the result of incomplete line opacity in our model stellar atmospheres.

The mass and temperature we find for LSS~3184 is consistent with those
expected based on the pulsation models of Monta\~n\'ez Rodr\'{\i}guez \&
Jeffery (\cite{mj02}) and with those found using optical photometry.
The discrepancy between the luminosity expected from the models and that
calculated based on UV observations remains a problem.

The chemical composition of LSS~3184 corresponds well to that expected from
a white dwarf merger. However, its mass is smaller than the merger models 
currently allow.

\begin{acknowledgements}
We acknowledge financial support from the Northern Ireland Department of
Culture, Arts, and Leisure and the UK PPARC (grant Ref
PPA/G/S/1998/00019).
\end{acknowledgements}

\end{document}